# QCD EXPLORER BASED eA AND γA COLLIDERS

H. Karadeniz, E. Recepoğlu, Sarayköy Nuclear Research and Training Center, Ankara, Turkey
S. Sultansoy, TOBB University of Economics and Technology, Ankara, Turkey

*Abstract*

TeV scale lepton-hadron and photon-hadron colliders are necessary both to clarify fundamental aspects of strong interactions and for adequate interpretation of the LHC data. Today, there are two realistic proposals for the post-HERA era, namely, QCD Explorer (QCD-E) and Large Hadron electron Collider (LHeC). Both QCD-E and LHeC can operate as eA colliders, whereas γp and γA options are unique for QCD-E. Another advantage of QCD-E is the possibility to increase the center of mass energy by lengthening of electron linac. In this presentation main parameters of the QCD-E nucleus options are discussed.

## INTRODUCTION

It is known that lepton-hadron collisions have been playing a crucial role in exploring the structure of the proton. For example, the structure of the proton, as well as the quark-parton model was resulted from investigation of high energy electron-nucleon scattering. The investigation of physics phenomena at extreme small x but sufficiently high $Q^2$ is very important for understanding the nature of strong interactions at all levels from nucleus to partons.

Today, linac-ring type machines seem to be the sole way to TeV scale in lepton-hadron collisions at the constituent level (see reviews [1-8] and references therein). Construction of future linear collider or a special e-linac tangentially to existing (HERA, TEVATRON, RHIC, LHC) or planned (VLHC) hadron rings will provide a number of new powerful tools in addition to ep and eA options:

- TeV scale γp [9, 10] and γA [11] colliders.
- FEL-Nucleus colliders [12, 13].

At the same time, the results from lepton-hadron colliders are necessary for adequate interpretation of physics at future hadron colliders. A $\sqrt{s} \approx 1$ TeV ep and γp colliders will be very useful in 2010's when precision era at LHC will begin. Approximately 10 years ago the THERA (TESLA on HERA) collider proposal was developed with this aim [14].

Today, there are two realistic proposals namely, QCD Explorer [6-8, 15-17] and LHeC [18]. During the last year ring-ring type LHeC is permanently transformed into QCD-E (see LHeC group's presentations at this conference [19-21]).

It should be noted that, the rational requirement of minimal influence on the LHC infrastructure excludes ring-ring option of the LHeC as well as super-bunch option of the QCD Explorer proposals. Therefore, sole reasonable choice for LHC based lepton-hadron collider is to construct special linac tangential to LHC ring.

## eA COLLIDER

Officially approved LHC ion programme contains an operation with Lead (Pb) ion beams and, the extension to lighter ions Tin (Sn), Krypton (Kr), Argon (Ar) and Oxygen (O) is under consideration [22].

The most transparent expression (in practical units) for the luminosity of linac-ring type eA colliders can be obtained from corresponding expression for ep colliders [23] with obvious modifications:

$$L = 4.8 \cdot 10^{30} \text{cm}^{-2}\text{s}^{-1} \cdot (n_A/10^{11}) \cdot (10^{-6}\text{m}/\varepsilon_A) \cdot (\gamma_A/1066)$$
$$\cdot (10\text{cm}/\beta_A) \cdot (P_e/22.6\text{MW}) \cdot (250\text{GeV}/E_e)$$

where $P_e$ denotes electron beam power, which is taken equal to radiation power of corresponding e-ring. With $E_e = 70$ GeV, $P_e = 50$ MW and LHC lead beam nominal parameters given in the Table 1, one obtains $L_{ePb} = 1.0 \cdot 10^{28}$ cm$^{-2}$s$^{-1}$. Let us mention that this value corresponds to nominal ion beam parameters together with ideal (special, dedicated) electron linac.

Table 1. Main parameters of the LHC-Pb beams [22]

|  | Nominal |
|---|---|
| Energy per charge, TeV | 7 |
| Transv. norm. emitt., $\varepsilon^*$, μm | 1.5 |
| β at the IP (coll.) $\beta^*$, m | 0.5 |
| r.m.s. beam radius at IP, $\sigma^*$, μm | 15 |
| Bunch spacing, $\tau_b$, ns | 124.75 |
| Number of bunches per ring k | 608 |
| Filling time per ring, min. | 9.8 |
| Number of ions per bunch; $N_b$ | 6.8x10$^7$ |
| IBS growth time (coll.), $\tau_\varepsilon$, h | 15 |

In the Table 2 we present parameters of LHC ion beams and eA luminosity estimations. In the last row results of our previous study for ep option are given [24, 25]. The values in parentheses correspond to upgraded LHC ion beams: $n_A \rightarrow 5n_A$, $\beta_A \rightarrow 0.1$m (in this case $\tau_{IBS} \approx 2$h).

Table 2. Nominal bunch intensities, IBS lifetimes [22] and eA luminosities for different ions in the LHC and ideal e-linac.

| Ion | $n_A$ | $\tau_{IBS}$, h | $L_{eA}$, cm$^{-2}$ m$^{-1}$ |
|---|---|---|---|
| Pb$^{82}_{208}$ | 6.8 x 10$^7$ | 15 | 1.0 x 10$^{28}$ (2.5 x 10$^{29}$) |
| Sn$^{50}_{120}$ | 2.8 x 10$^8$ | 10 | 4.1 x 10$^{28}$ (1.0 x 10$^{30}$) |
| Kr$^{36}_{84}$ | 5.5 x 10$^8$ | 10 | 8.3 x 10$^{28}$ (2.1 x 10$^{30}$) |
| Ar$^{18}_{40}$ | 2.2 x 10$^9$ | 10 | 3.5 x 10$^{29}$ (8.7 x 10$^{30}$) |
| O$^8_{16}$ | 1.2 x 10$^{10}$ | 10 | 2.1 x 10$^{30}$ (5.2 x 10$^{31}$) |
| p | 1.7 x 10$^{11}$ | 10 | 2.4 x 10$^{31}$ (6.0 x 10$^{32}$) |

Concerning the actual linear collider proposals, as it is shown in [24, 25] for ep option, in order to achieve sufficiently high luminosities the CLIC technology should be radically modified, whereas moderate modifications will be enough in the case of TESLA. Nominal parameters of the TESLA and CLIC beams are presented in the Table 3. The main advantage of the CLIC technology is higher accelerating gradient and, therefore, shorter linac length. It is seen that nominal TESLA bunch spacing is close to that of the LHC ion beam.

Table 3. Nominal parameters of the TESLA and CLIC

|  | TESLA | CLIC |
| --- | --- | --- |
| Accelerating gradient, MeV/m | 23.4 | 150 |
| Bunch spacing $\tau_e$, ns | 200 | 0.66 |
| Number of bunches per pulse $n_b$ | 5600 | 154 |
| Repetition rate $f_{rep}$, Hz | 5 | 200 |
| Electrons per bunch $n_e$, $10^{10}$ | 2 | 0.4 |

In the Table 4 we present maximum number of electrons in a bunch (requiring $\Delta Q_A < 0.01$) and achievable luminosities (assuming $n_A \to 5 n_A$, $\beta_A \to 0.1$ m and $\tau_A = \tau_e = 100$ ns) for different ion species.

Table 4. Maximum e-bunch intensities and eA luminosities for different ions in the LHC and TESLA-like e-linac.

| Ion | $n_e$ | $L_{eA}$, cm$^{-2}$ m$^{-1}$ | $A \times L_{eA}$, cm$^{-2}$ m$^{-1}$ |
| --- | --- | --- | --- |
| $Pb^{82}_{208}$ | 1.0 x 10$^9$ | 1.0 x 10$^{27}$ | 2.1 x 10$^{29}$ |
| $Sn^{50}_{120}$ | 1.6 x 10$^9$ | 6.6 x 10$^{27}$ | 7.9 x 10$^{29}$ |
| $Kr^{36}_{84}$ | 2.2 x 10$^9$ | 1.8 x 10$^{28}$ | 1.5 x 10$^{30}$ |
| $Ar^{18}_{40}$ | 4.4 x 10$^9$ | 1.4 x 10$^{29}$ | 5.6 x 10$^{30}$ |
| $O^{8}_{16}$ | 1.0 x 10$^{10}$ | 1.8 x 10$^{30}$ | 2.9 x 10$^{31}$ |
| p | 8.0 x 10$^{10}$ | 5.0 x 10$^{31}$ | 5.0 x 10$^{31}$ |

Let us note that luminosity values given in the Table 4 can be increased by a factor 10 or even more with energy recovery linac [26].

## γA COLLIDER

Luminosity of γA collisions practically coincides with luminosity of eA collisions for small ($l \ll 1$m) distance between conversion region and interaction point (conversion efficiency factor 0.6 could be compensated due to smaller transverse size of photon beam comparing to nucleus beam). The luminosity will be reduced by a factor 2÷3 at $l = 3$m [10]. Opposite polarization of electron and laser photon beams will be advantageous both for higher luminosity and better monochromaticity of high energy γ beam at interaction point. These topics are under study.

## PHYSICS GOALS

Obviously, the physics search program for eA option coincides with that of the LHeC [18]. A preliminary list of physics goals of the QCD Explorer based γA collider comprises [4, 10, 14]:

- Total cross-section to clarify real mechanism of very high energy γ-nucleus interactions.
- Investigation of a hadronic structure of the photon in nuclear medium.
- According to the vector meson dominance (VMD) model the proposed machine will also be a ρ-nucleus collider. This will result in formation of quark-gluon plasma at very high temperature but relatively low nuclear density.
- The gluon distribution at extremely small $x_g$ in nuclear medium ($\gamma A \to Q\bar{Q} + X$; Q = c, b).
- Investigation of both heavy quarkonia and nuclear medium properties ($\gamma A \to J/\Psi(Y) + X, J/\Psi(Y) \to l^+ l^-$)
- Existence of multi-quark clusters in nuclear medium and a few-nucleon correlation.

γA collider will give unique opportunity to investigate the small $x_g$ region in nuclear medium[27]. Indeed, due to the advantage of the real γ spectrum, heavy quarks will be produced via γg fusion at a characteristic x parameter,

$$x_g \approx \frac{5 m_{c(b)}^2}{0.8 \times (Z/A) \times s_{ep}}$$

which is approximately (2÷3)x10$^{-5}$ for charmed and (2÷3)x10$^{-4}$ for beauty hadrons. The numbers of $c\bar{c}$ and $b\bar{b}$ pairs, which will be produced in γA collisions, can be estimated as $10^6 \div 10^7$ and $10^5 \div 10^6$ per working year, respectively. Therefore, one will be able to investigate the small $x_g$ region in detail. For this reason, a very forward detector in γ-beam direction will be useful for investigation of small $x_g$ region via detection of charmed and beauty hadrons.

## CONCLUSION

Lepton-hadron collider with $\sqrt{s_{ep}} > 1$ TeV is necessary both to clarify fundamental aspects of the QCD part of the Standard Model and for adequate interpretation of experimental data from the LHC. Today, there are two realistic proposals, namely, QCD Explorer and LHeC. Both QCD-E and LHeC will give opportunity to achieve sufficiently high luminosity to explore crucial aspects of the strong interactions. Whereas LHeC is based on the more familiar approach (we have nice experience from the HERA), QCD-E has a number of advantages:

- additional γp, γA and FEL γA options
- electron beam energy can be expanded by increasing linac length, whereas synchrotron radiation blocks this road for LHeC
- minimal influence on the LHC tunnel.

The main goal of both QCD-E and LHeC proposals is to clarify fundamental aspects of strong interactions. Their potential for the BSM physics search is restricted by center of mass energy. Therefore, very high luminosity is not so important. In our opinion γA option of the QCD-E will provide crucial information on QCD dynamics at small $x_g$ in nuclear medium

This work is partially supported by the Turkish State Planning Organization under the grant No DPT2002K-120250 and Turkish Atomic Energy Authority.